\documentclass[%
 aip,
 amsmath,amssymb,
 reprint,%
,floatfix]{revtex4-1}

\usepackage{graphicx}
\usepackage{subcaption}
\usepackage{dcolumn}
\usepackage{bm}
\usepackage{caption}
\usepackage{siunitx}
\usepackage{rotating}   
\usepackage{placeins}
\usepackage{booktabs}
\let\oldbibliography\bibliography
\renewcommand{\bibliography}[1]{\FloatBarrier\oldbibliography{#1}}

\usepackage[utf8]{inputenc}
\usepackage[T1]{fontenc}
\usepackage{newunicodechar}
\newunicodechar{，}{,}
\usepackage{mathptmx}
\usepackage{etoolbox}
\usepackage{float}
\usepackage{hyperref}   
\usepackage{xcolor}
\usepackage[normalem]{ulem} 
\makeatletter
\def\@email#1#2{%
 \endgroup
 \patchcmd{\titleblock@produce}
  {\frontmatter@RRAPformat}
  {\frontmatter@RRAPformat{\produce@RRAP{*#1\href{mailto:#2}{#2}}}\frontmatter@RRAPformat}
  {}{}
}%
\makeatother
\begin{document}


\title{Development of a High-Performance Permanent Magnet System for Ion Trapping Experiments}

\author{Jifei Wu}
\affiliation{ 
Institute of Modern Physics, Fudan University, Shanghai 200433, China
}%
\affiliation{%
Key Laboratory of Nuclear Physics and Ion-Beam Application (MOE), Fudan University, Shanghai 200433, China
}%

\author{Jiawei Wang}
\affiliation{ 
Institute of Modern Physics, Fudan University, Shanghai 200433, China
}%
\affiliation{%
Key Laboratory of Nuclear Physics and Ion-Beam Application (MOE), Fudan University, Shanghai 200433, China
}%

\author{Tianhang Zhang}
\affiliation{ 
Institute of Modern Physics, Fudan University, Shanghai 200433, China
}%
\affiliation{%
Key Laboratory of Nuclear Physics and Ion-Beam Application (MOE), Fudan University, Shanghai 200433, China
}%

\author{Zichen Su}
\affiliation{ 
Institute of Modern Physics, Fudan University, Shanghai 200433, China
}%
\affiliation{%
Key Laboratory of Nuclear Physics and Ion-Beam Application (MOE), Fudan University, Shanghai 200433, China
}%

\author{Liangyu Huang}
\email{liangyuhuang@fudan.edu.cn}
\affiliation{ 
Institute of Modern Physics, Fudan University, Shanghai 200433, China
}%
\affiliation{%
Key Laboratory of Nuclear Physics and Ion-Beam Application (MOE), Fudan University, Shanghai 200433, China
}%

\author{Wei Wu}
\affiliation{Shanghai APACTRON Particle Equipment Co.,Ltd}%

\author{Bingsheng Tu}
\email{bingshengtu@fudan.edu.cn}
\affiliation{ 
Institute of Modern Physics, Fudan University, Shanghai 200433, China
}%
\affiliation{%
Key Laboratory of Nuclear Physics and Ion-Beam Application (MOE), Fudan University, Shanghai 200433, China
}%


\begin{abstract}
This work presents the design and fabrication of a compact permanent magnet based on an optimized stacked structure of fifteen NdFeB rings. The tunable NS‑SN‑NS configuration generates a central magnetic field of 0.8\,T with a reconstructed uniformity of 99.988\% within a 1\,mm radius spherical volume. The remaining field inhomogeneity is dominated by radial dipole components. Requiring neither cryogenics nor external power, this design provides a high‑performance and cost‑effective alternative to superconducting magnets for applications in ion‑trap development and Fourier‑transform ion cyclotron resonance mass spectrometry.
\end{abstract}

\maketitle

\section{\label{sec:level1}Introduction}

Modern atomic and molecular physics experiments across diverse fields—from precision spectroscopy and mass measurements to quantum information science—fundamentally rely on highly controlled magnetic field environments. Critical facilities such as Electron Beam Ion Traps (EBITs) \cite{Martnez2007TheHE,shlyaptseva2003x,nakamura2010activities}, Penning traps \cite{repp2012pentatrap,sturm2019alphatrap,smorra2015base,schwarz2003low,dilling2006mass,ding2020lorentz,wang2025development,zhang2025}, and Fourier Transform Ion Cyclotron Resonance mass spectrometers (FT-ICR MS) \cite{smith201821} impose stringent requirements on field strength, stability, spatial homogeneity, and the size of the usable field region. For instance, EBITs utilize strong fields (exceeding 7 T in some cases) to compress electron beams for producing highly charged ions \cite{Martnez2007TheHE}, while ultra-high-precision mass measurements in Penning traps and FT-ICR MS demand exceptionally homogeneous and stable fields to achieve sub-ppm resolution \cite{tu2011precision, repp2012pentatrap, smith201821}.

These requirements are traditionally met by three principal technologies. Superconducting magnets offer the highest field strengths and superior stability \cite{smith201821, Fawley2025, Künstner2024}, but their high cost, large footprint, and reliance on complex cryogenic systems pose significant barriers for many applications. Electromagnets offer flexible tunability; however, they require continuous high power and active cooling, which introduce challenges associated with thermal management, operational costs, and electromagnetic noise \cite{Wen2020, Luna2025}. In contrast, permanent magnets present a compelling alternative with inherent advantages: they require no electrical power during operation, generate minimal heat, and offer reasonable long-term stability in a compact and potentially cost-effective package \cite{Barnes2020, Adambukulam2020}.

Consequently, advanced permanent magnet systems have attracted considerable interest as viable, high-performance alternatives. Recent research demonstrates significant progress in overcoming historical limitations of field strength and homogeneity. Compact Penning traps utilizing NdFeB magnets have been realized for applications like laser cooling of ions and electron beam confinement, generating stable fields of ~0.6–0.7 T \cite{McMahon2019, Barnes2020, Suess2002}. Innovative design strategies have focused on optimizing field characteristics. Topology optimization has been applied to enhance Halbach cylinder efficiency \cite{Bjork2016}, while non-circular array designs have improved homogeneity by orders of magnitude \cite{Tewari2023}. Tunable systems have also been developed, ranging from adjustable ring magnets providing 0.18–0.27 T fields \cite{Pizzey2021} to mechanically rotatable structures for precise field control \cite{Berzins2021}. For extreme stability requirements, such as in quantum information experiments, Halbach arrays combined with soft magnetic materials have achieved 1.5 T fields with drifts below 3 ppb/h at cryogenic temperatures \cite{Adambukulam2020}, similar principles have been successfully adapted for MRI applications \cite{Tadic2011,trenec2011permanent}. However, the system is relatively large as the magnet alone weighs over 60~kg. For FT‑ICR mass spectrometry applications, Lemaire et al.~\cite{lemaire2018compact} developed a compact spectrometer based on a Halbach‑type permanent magnet, achieving a central field of about 1 T with a relative homogeneity of $0.5\%$ over a 3 cm cube.

High‑precision FT‑ICR mass spectrometry or advanced ion‑trap experiments demand magnetic fields with extremely high local uniformity on the millimetre scale, while still keeping the overall size and weight moderate. The goal of this work is therefore to develop a compact permanent‑magnet system that achieves improved local uniformity and a reasonable central field without increasing the physical dimensions or weight, making it suitable for both FT‑ICR and ion‑trap applications.

To meet this challenge, we propose a novel NS‑SN‑NS stacked configuration of fifteen NdFeB ring magnets. This design allows independent adjustment of inter‑group gaps, enabling active compensation of field asymmetries – a feature absent in fixed Halbach assemblies. The fabricated magnet generates a central field of approximately 800 mT with an inner bore diameter of 44 mm, achieves a relative homogeneity of 0.063\% over a 1 cm sphere, which could be even better after the optimization of the field symmetry. The remaining of this paper is organized as follows: in Section~\ref{secII} , we present the principle of the magnet design supported by the functionality of the hardware and the magnetic field simulation. Section~\ref{secIII} reports the measured magnetic-field strength distributions at different stages of the assembly process and discusses the optimization of field homogeneity. Finally, conclusions and an outlook are given in Section~\ref{secIV}.

\begin{figure*}[!htbp]
\includegraphics[width=1\textwidth]{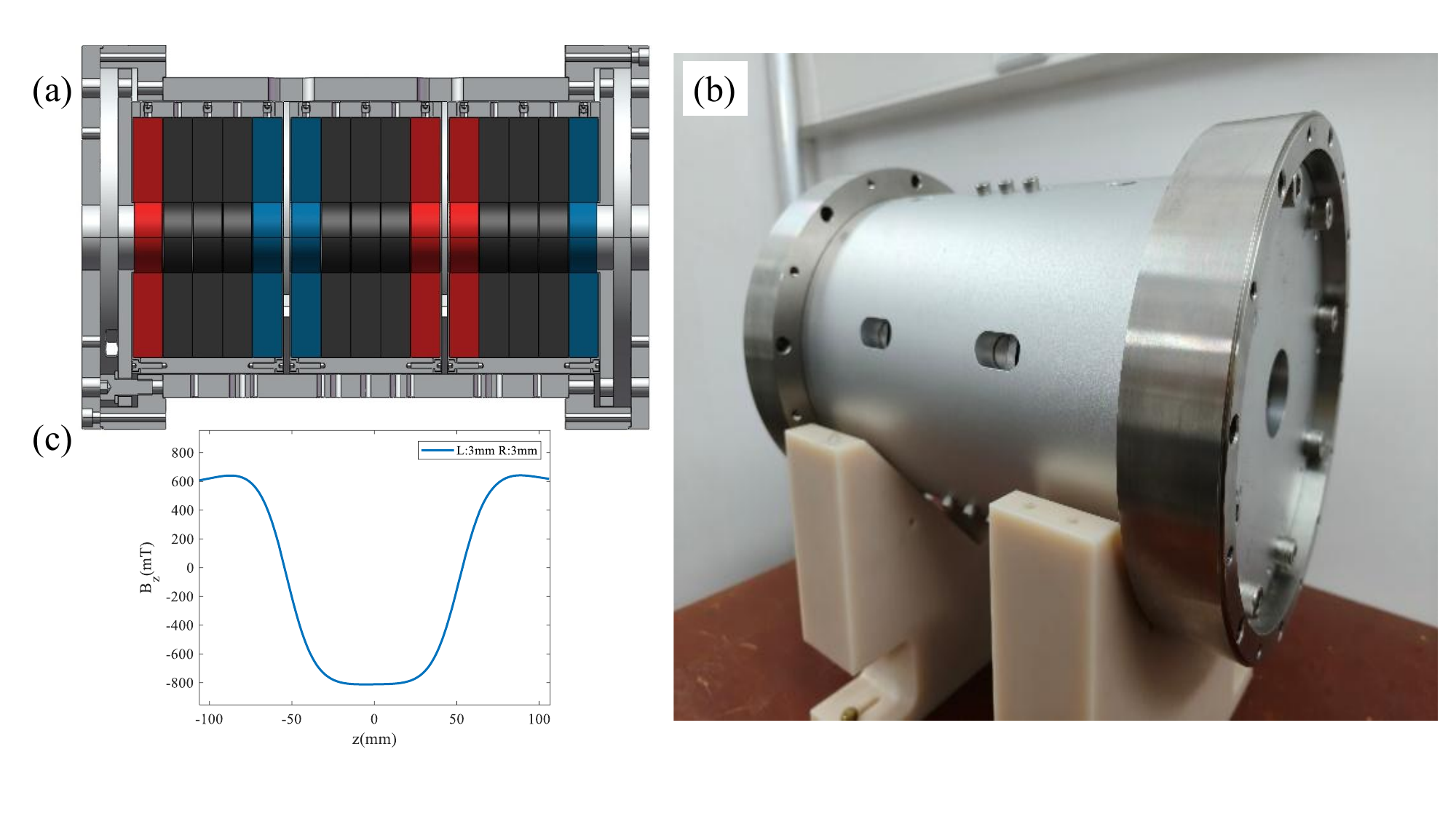} %
\caption{\label{fig:magassembly} Schematic illustration of the magnet design and assembly. (a) Conceptual structure showing fifteen NdFeB ring magnets (outer diameter $150$~mm, inner diameter $44$~mm, thickness $10$~mm each) arranged into three groups of five. The left, center, and right groups are labelled L, O, and R, respectively. Red indicates north pole, blue indicates south pole. The $z$‑axis direction is from L through O to R. (b) Photograph of the physical prototype, where three five‑magnet groups are encapsulated in an aluminum sleeve and mounted inside a larger aluminum housing with adjustable inter‑group spacing. (c) FEM simulation result of the axial magnetic field distribution, showing field compression at the geometric center. The distances between groups are indicated: $\Delta_{\mathrm{L-O}}=3$~mm and $\Delta_{\mathrm{O-R}}=3$~mm. }
\end{figure*}

\section{Principle and Design}\label{secII}
The magnet assembly consists of fifteen neodymium--iron--boron (NdFeB, grade N42) ring magnets, with a unit cost of less than \num{1000} RMB. Each magnet has an inner diameter of \SI{44}{mm}, an outer diameter of \SI{150}{mm}, and a thickness of 10 mm. The N42 grade was selected as it offers an optimal trade-off between remanence, coercivity, cost-effectiveness, and widespread commercial availability, making it well suited for constructing compact permanent-magnet systems of this scale. The employed N42 NdFeB magnets exhibit a temperature coefficient of remanence of approximately \SI{-0.12}{\percent\per\degreeCelsius}\cite{QuadrantN42Datasheet,Haavisto2009}. As illustrated in Fig.~\ref{fig:magassembly}(a), fifteen magnets are divided into 3 groups, each comprising five magnets that are attracted with each other and stick together. The magnetic field lines of the middle group of magnets are oriented in the opposite direction to those of the two outer groups. In this configuration, the magnetic field at the center is significantly compressed, resulting in a highly concentrated field. The homogeneity of the magnetic field strongly depends on the geometry; therefore, a finite element method (FEM) was employed to simulate the magnetic field and optimize the geometry. The FEM static magnetic field module solves the following equations, derived from Maxwell's equations and incorporating a relative permeability model, known as $\triangledown$B=0 and B=$\mu_0$$\mu_r$H. The result shown in Fig.~\ref{fig:magassembly}(c) demonstrates that the simulated magnetic field strength is approximately 0.8 T and the homogeneity is discussed in Section III. 

The physical assembly of the magnets is shown in Fig.~\ref{fig:magassembly}(b). Each group of five assembled pieces of magnets is encapsulated within an aluminum alloy sleeve featuring central hollows on the left and right surfaces. Subsequently, a homogeneous axial magnetic field, similar to the field of a solenoid, was produced using this assembly. In principle, this assembled group of magnets could be replaced by a single, larger magnet. However, fabricating a magnet of this size as a single integral piece is highly challenging. Owing to the strong magnetic field intensity, relying solely on magnetic attraction for assembly entails the risk of violent collisions, which could lead to damage or even fragmentation of the magnet. During this assembly process, the repulsive forces between the three magnet groups were strategically utilized to facilitate their assembly. Specifically, we designed a fixture to align the central holes of the three magnet groups, after which pressure was gradually applied from both sides to press the left and right groups toward the center, and the assembly was secured using set screws. The repulsive force between the magnet groups was estimated to be approximately 8 kN, underscoring the necessity of the specially designed fixture and controlled pressing process to avoid violent collisions. 

The magnetization process of permanent magnets inevitably introduces non-uniformity or asymmetry. By pre-measuring the individual pieces of magnets and arranging the five magnets in a proper configuration, a magnetic field distribution matching the FEM simulation can be achieved. In addition, a tunable space with a range of 3~mm to 7~mm was maintained between the three magnet groups using a set of tightness screws. This enables fine-tuning of the axial magnetic field symmetry. Note that insufficient spacing may lead to magnet demagnetization. Details of the homogeneity adjustment procedure are discussed in the Section~\ref{secIII}. 

\section{Result and Discussion}\label{secIII}
The magnetic field components $B_x$, $B_y$, and $B_z$ were measured using a high-precision 3D magnetic field measurement instrument (SENIS Teslameter 3MH6-E) with an interchangeable Hall probe of Type C~\cite{Senis2023}. The probe has a sensitive volume for the magnetic field of $0.1$\,mm $\times$ $0.01$\,mm $\times$ $0.1$\,mm, providing high spatial resolution, while its housing dimensions are $4.0$\,mm $\times$ $0.9$\,mm $\times$ $8.0$\,mm. The probe was mounted on an axial guide rail via an acrylic holder. The teslameter has a DC accuracy of $\pm0.01\%$ of the selected range and a resolution of $1.2$\,\textmu T rms at a sampling rate of 100 SPS, and the axial guide rail provides a positioning accuracy of $\pm0.1$\,mm.

\begin{figure}[!htbp]
 \begin{subfigure}{0.45\textwidth}
\includegraphics[width=\linewidth]{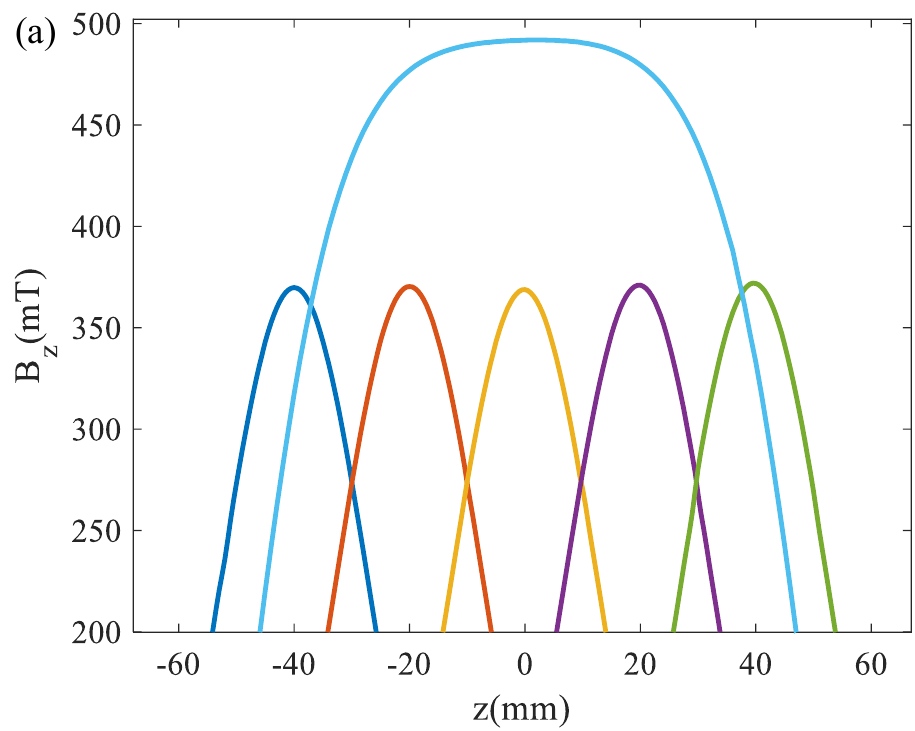} %
\caption*{}
\end{subfigure}
\hfill
\begin{subfigure}{0.45\textwidth}
\includegraphics[width=\linewidth]{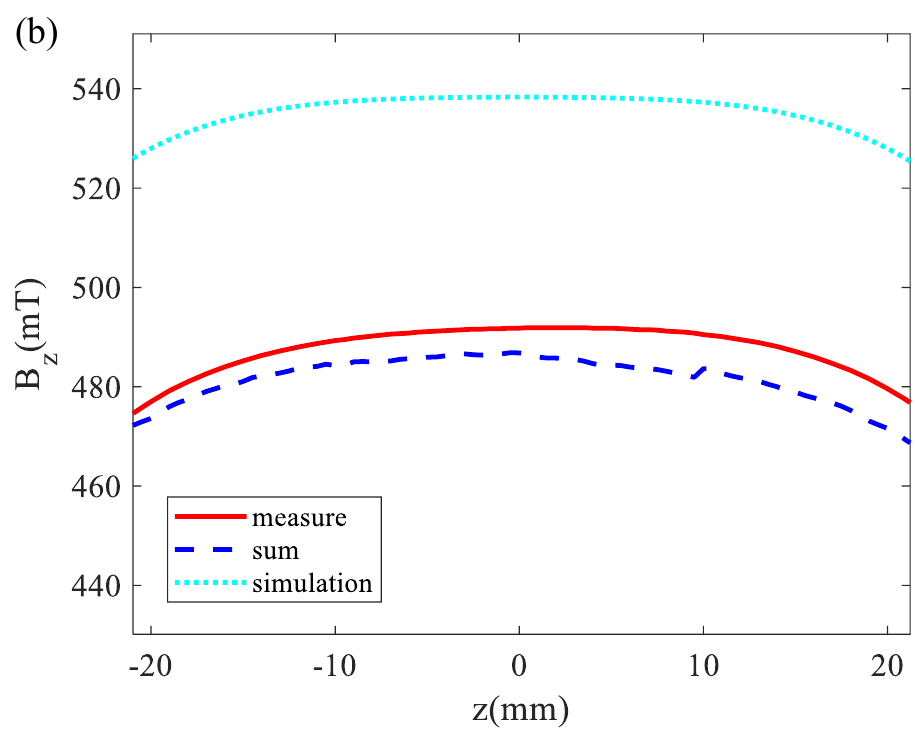} %
\caption*{}
\end{subfigure}
\caption{(a) Measured axial field distributions of individual magnets and the total field of an assembled five‑magnet group. (b) Comparison of the measured axial field distribution (red solid line), FEM simulation (cyan dash line), and the sum result (blue dot line) for five individual magnets within $-20\le z\le20$~mm. The small kink visible in the sum curve is not solely caused by the systematic accuracy of the teslameter (0.01\%) or the positioning stage (0.1\,mm); it arises from the accumulation of random errors during the separate measurements of the five individual magnets. These errors include random origin shifts (up to $\pm0.1$\,mm) due to positioning repeatability and random reading fluctuations within the teslameter resolution. When the five curves are summed point‑wise, these random deviations accumulate and produce the observed kink. In the physically assembled group, the magnets share a common reference and the random reading errors are averaged out, so the kink disappears. } 
\label{fig:five_magnet_total} 
\end{figure}

Magnetic field measurement starts with a two‑step calibration: xy‑plane center calibration and z‑axis calibration. First, the axial field of each individual ring magnet was measured. From a collection of magnets sourced from multiple commercial batches, five magnets with closely matched peak fields were selected to form the central group (denoted as group O, for “center”). Their individual axial profiles and the assembled group O field are shown in Fig.~\ref{fig:five_magnet_total}(a). The peak fields of the five single magnets are 369.8 mT, 370.3 mT, 368.8 mT, 371.0 mT, and 372.0 mT, respectively. By arranging them symmetrically according to their peak values, the assembled group O achieves a uniform central region with a peak field of 492.2 mT at $z=0$ (Fig.~\ref{fig:five_magnet_total}b). The “sum” curve in Fig.~\ref{fig:five_magnet_total}(b) was obtained by point‑wise addition of the five individual measurements, confirming linear superposition. The original measurements used a variable step size (0.5 mm near the center, increasing to 1 mm and then 2 mm at larger $|z|$). To enable a point‑wise summation, the data were resampled onto a uniform grid of 0.5 mm using cubic spline interpolation before summation.

\begin{figure}[!htbp]
\centering
\includegraphics[width=0.45\textwidth]{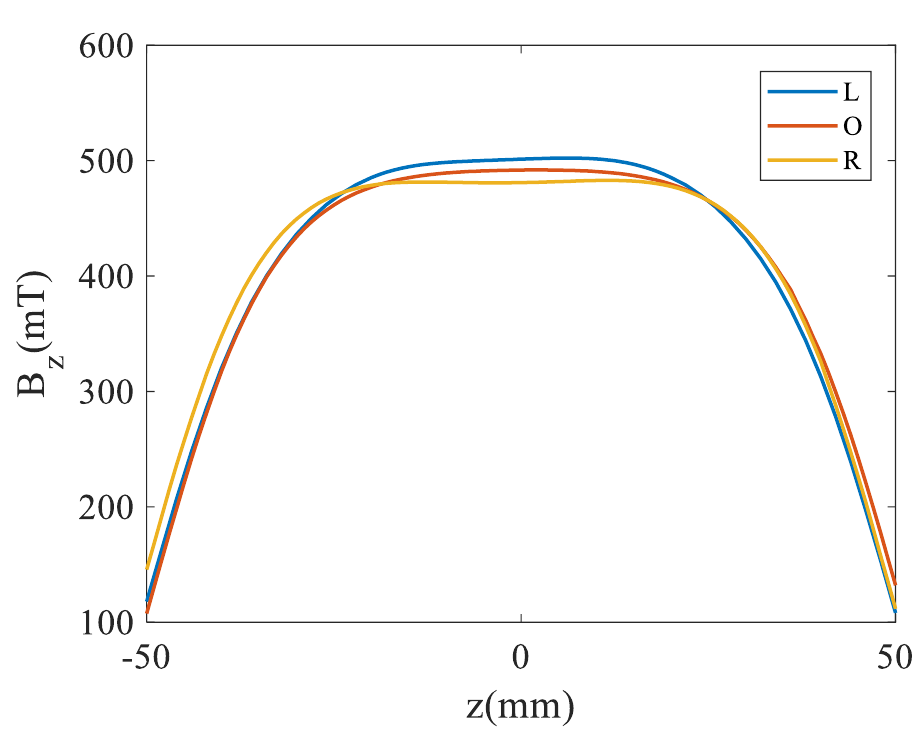}
\caption{Axial magnetic field profiles of the three five‑magnet groups L (left), O (center), and R (right) measured along the $z$‑axis. Group O exhibits the flattest and most symmetric central region, while groups L and R show different peak fields and poorer symmetry due to batch‑to‑batch variation.}
\label{fig:group_profiles}
\end{figure}

Two additional five‑magnet groups, labeled L (left) and R (right), were assembled using magnets from different batches. As a result, their axial profiles exhibit not only different peak fields but also poorer symmetry compared to group O. As shown in Fig.~\ref{fig:group_profiles}, group L has a peak of 503.0 mT at $z=10$ mm, and group R peaks at 483.5 mT at $z=-10$mm. Their profiles are visibly less symmetric than that of group O. Because group O possesses the flattest and most symmetric central region, it was deliberately placed at the center of the final assembly, where its characteristics dominate the field in the ion‑trap region.

\begin{figure}[!htbp]
\includegraphics[width=0.45\textwidth]{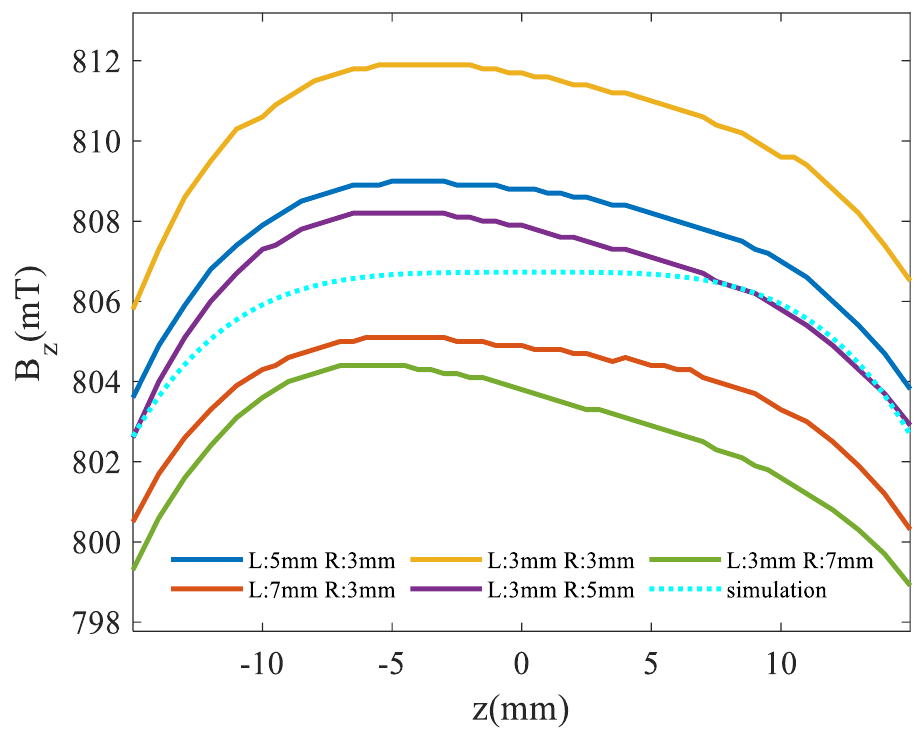} %
\caption{\label{fig：symmetry} Measured (solid line) and simulated (dash line) axial magnetic field distributions at $z = \pm15$~mm for different spacing configurations between the three magnet groups (L and R denote the spacing between the center magnet and the left and the right magnets, respectively).}
\end{figure}

The complete magnet assembly consists of these three groups (left, center, right) stacked coaxially with the central group magnetized opposite to the outer groups, which compresses the field at the geometric center and produces a central field above 800 mT (Fig.~\ref{fig：symmetry}). The axial field measurement generally agrees with the FEM simulation. However, an asymmetry in $B_z$ is observed in the central region from $-10$mm to $10$mm (Fig.~\ref{fig：symmetry}), mainly because the intrinsic differences among groups L, O, and R introduce a gradient along the $z$‑axis. To compensate this asymmetry, we adjusted the gaps between the groups. By systematically varying the left and right spacings within the mechanically limited range of 3--7~mm (see Section~II), we found that setting the left gap to 7~mm and the right gap to 3~mm minimizes the asymmetry while maintaining a high central field. The resulting field profile is shown in Fig.~\ref{fig：symmetry} (red solid line). These gap settings were found to give the best homogeneity achievable under the mechanical constraints of the current assembly.
The residual asymmetry in the $B_z$ profile is attributed to the inherent mismatch between the L and R groups, which came from different production batches and consequently exhibit different peak fields and symmetry. A better strategy for future designs would be to select two groups with closely matched but mirrored magnetic field profiles, which would minimize the intrinsic gradient and further improve the achievable homogeneity.

The magnetic axis deviation was determined through a two-step measurement procedure. First, on both end-faces of the cylindrical magnet, the radial component of the magnetic field was measured along a circular trajectory with a radius of 11 mm, centered on the geometric center. By fitting the radial field data as a function of the azimuthal angle, the angular position corresponding to the field minimum was identified, thereby determining the phase of the magnetic axis deviation. Subsequently, a linear scan was performed along this specific phase direction, covering a length of 22 mm centered at the geometric center. The zero-crossing point of the radial field component along this line was precisely located, determining the magnitude of the magnetic axis displacement from the geometric center.

\begin{figure}[!htbp]
 \begin{subfigure}{0.45\textwidth}
        \includegraphics[width=\linewidth]{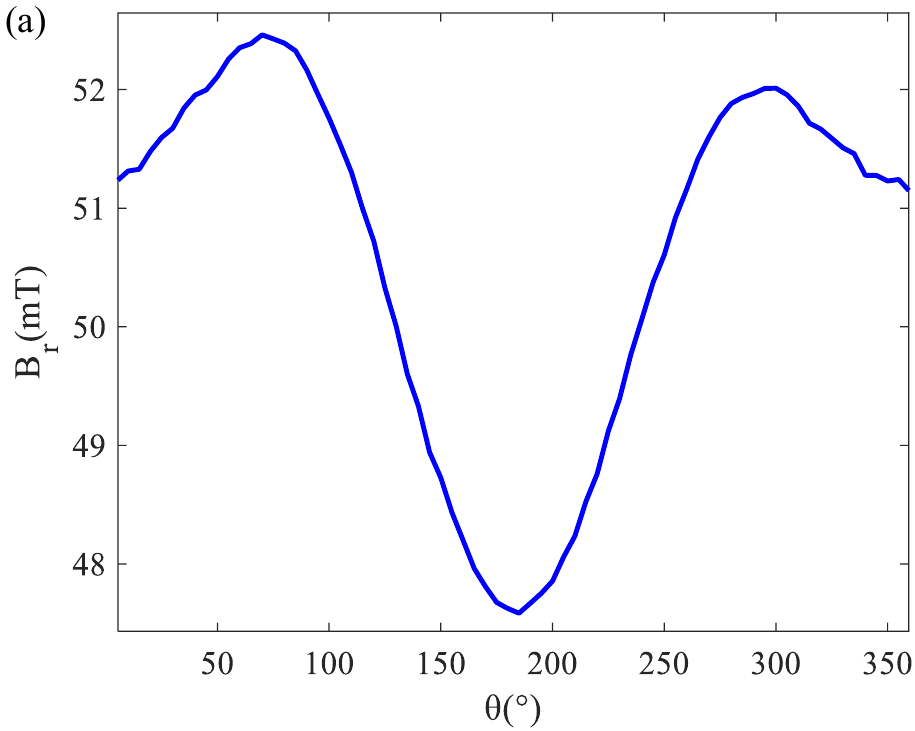}  
        \caption*{} 
    \end{subfigure}
    \hfill 
    \begin{subfigure}{0.45\textwidth}
        \includegraphics[width=\linewidth]{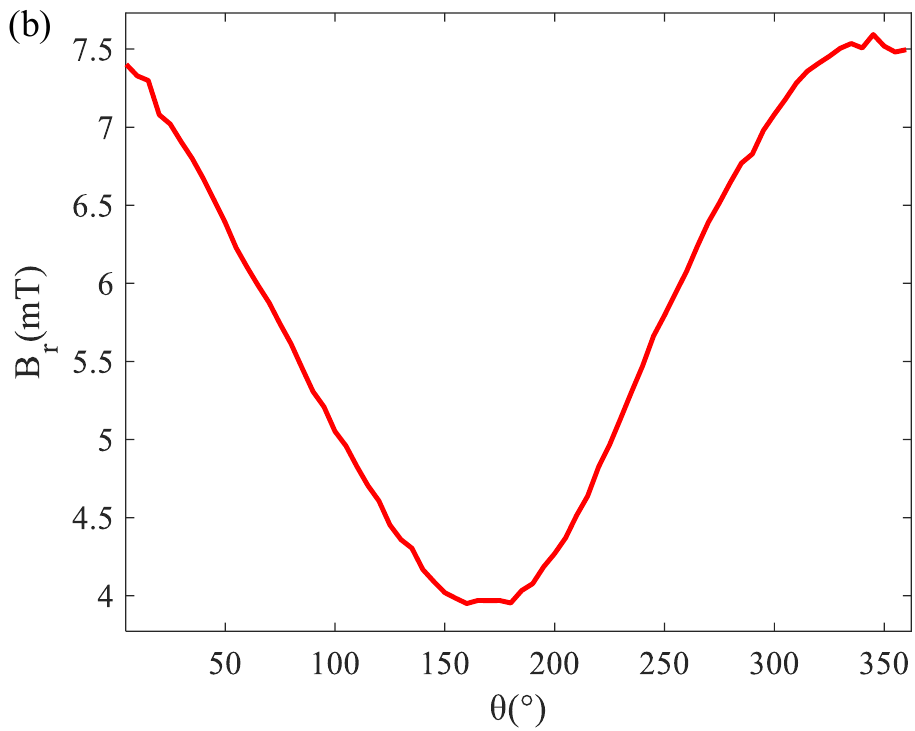}
        \caption*{}
     \end{subfigure}
\caption{Measurement of the radial magnetic field component $B_r$ as a function of azimuthal angle on both end‑faces of the magnet assembly. (a) Data acquired on the right end‑face; (b) Data acquired on the left end‑face. The difference between the two panels is due to the asymmetric spacing between the magnet groups ($7$~mm left, $3$~mm right). } 
\label{fig:magnet axis theta}
\end{figure}

\begin{figure}
\includegraphics[width=0.45\textwidth]{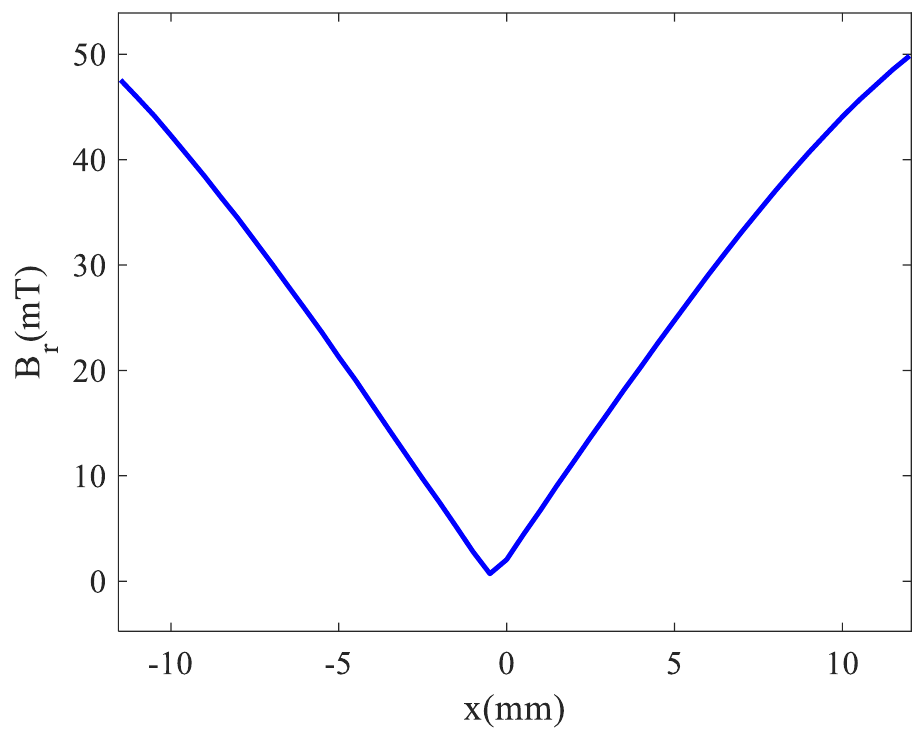} %
\caption{\label{fig:magnet axis x}Measurement of the radial magnetic field component $B_r$ along a diameter on the right end‑face of the magnet assembly. The scan direction corresponds to the phase determined from Fig.~4 (approximately $180^\circ$). The minimum field occurs at $x = -1$~mm, indicating a magnetic axis offset of $1$~mm relative to the geometric center (which is set as the reference point $x=0$).}
\end{figure}

Fig.~\ref{fig:magnet axis theta}(a) and (b) show the measured radial field component $B_r$ as a function of azimuthal angle on both end-faces of the magnet assembly. The linear scan of the radial field along the identified phase direction (approximately 180$^\circ$) is shown in Fig.~\ref{fig:magnet axis x}, revealing the displacement of the magnetic axis from the geometric center. From these measurements, we determined that the magnetic axis is offset by approximately 1.0 mm in the direction of 180$^\circ$ (i.e., along the negative $x$-axis) relative to the geometric center, accounting for approximately 2.3\% of the inner radius of the magnet assembly.

All performance data reported in this work, including the achieved field strength and homogeneity, were obtained prior to any final corrective alignment of the magnetic axis. Therefore, these results represent a conservative estimate of the system's performance capability. It is anticipated that the performance could be further enhanced to approach the ideal scenario through straightforward mechanical alignment of the ion trap axis to the magnetic axis.

Then we present a comprehensive methodology for magnetic field uniformity analysis, integrating experimental measurement with advanced mathematical modeling. Magnetic field data were acquired using a three-axis Hall sensor system at maximum field position， systematically scanning the region of interest with particular focus on two concentric spherical surfaces of 5.0 mm and 10.0 mm radii. Following standard measurement procedures including coordinate transformation to spherical coordinates, the axial field component ($B_z$) was selected as the primary parameter for uniformity characterization.

The magnetic field in the source‑free central region can be described by a scalar potential $\Phi$ satisfying Laplace’s equation. In spherical coordinates $(r,\theta,\phi)$ with origin at the geometric center, the potential is expanded in real spherical harmonics as
\[
\Phi(r,\theta,\phi)=\sum_{l=1}^{l_{\max}}\sum_{m=-l}^{l}a_{lm}\left(\frac{r}{r_0}\right)^{l}Y_{lm}(\theta,\phi), \tag{1}
\]
where $r_0$ is a reference radius (taken as the mean radius of all measurement points, approximately $7.5$mm), $Y_{lm}$ are orthonormal real spherical harmonics, and $a_{lm}$ are the expansion coefficients (units T·m). The axial magnetic field component $B_z = -\partial\Phi/\partial z$ is then given by
\[
B_z(r,\theta,\phi)=-\sum_{l,m}a_{lm}\left(\frac{r}{r_{0}}\right)^{l}\left[\frac{l\cos\theta}{r}Y_{lm}-\frac{\sin\theta}{r}\frac{\partial Y_{lm}}{\partial\theta}\right]. \tag{2}
\]

For a set of $N$ measurement points $(r_i,\theta_i,\phi_i)$ with values $B_{z,i}$, we construct a linear system $\mathbf{A}\mathbf{a} = \mathbf{b}$, where $\mathbf{a}$ is the vector of coefficients $a_{lm}$, $\mathbf{b}$ contains the measured $B_{z,i}$, and the design matrix $\mathbf{A}$ has elements given by the bracketed expression multiplied by $-(r_i/r_0)^l$.

Performing singular value decomposition (SVD) on the design matrix $\mathbf{A}$ yields
\[
\mathbf{A} = \mathbf{U}\boldsymbol{\Sigma}\mathbf{V}^T, \tag{3}
\]
where $\boldsymbol{\Sigma} = \operatorname{diag}(\sigma_1,\sigma_2,\dots,\sigma_n)$ with $\sigma_1\ge\sigma_2\ge\cdots\ge\sigma_n\ge0$,
and $\mathbf{U}$, $\mathbf{V}$ are orthogonal matrices. The condition number of $\mathbf{A}$ is then given by
\[
\kappa(\mathbf{A}) = \frac{\sigma_1}{\sigma_n}. \tag{4}
\]
For $l_{\max}=5$, $\kappa \approx 3.3\times10^{11}$, indicating that the linear system $\mathbf{A}\mathbf{a}=\mathbf{b}$ is severely ill‑conditioned. Consequently, the solution $\mathbf{a}$ is highly sensitive to perturbations in $\mathbf{A}$ and $\mathbf{b}$, and measurement noise would be greatly amplified if a direct least‑squares method were used.

To overcome this difficulty, we employ truncated singular value decomposition (TSVD)~\cite{Golub2013, Hansen1998}. TSVD discards the smallest singular values that are dominated by numerical noise, retaining only those larger than $10^{-6}\sigma_1$. For $l_{\max}=5$, this keeps $25$ out of $35$ singular values. Remarkably, these 25 singular values correspond exactly to the number of physically observable multipole modes: among the 35 spherical harmonic coefficients, the 10 coefficients with $m=\pm l$ (e.g., $l=1,m=\pm1$, $l=2,m=\pm2$, …, $l=5,m=\pm5$) contribute nothing to $B_z$ because their angular dependence contains no $z$ component, making the corresponding columns of $\mathbf{A}$ identically zero. Consequently, the design matrix has an effective rank of 25, and the TSVD truncation naturally eliminates these unobservable modes, thereby improving numerical stability without losing any physically relevant information. The reference radius $r_0$ is chosen to keep the magnitudes of the basis functions balanced across all orders, further improving numerical stability.

\begin{figure}[!htbp]
 \begin{subfigure}{0.48\textwidth}
\includegraphics[width=\linewidth]{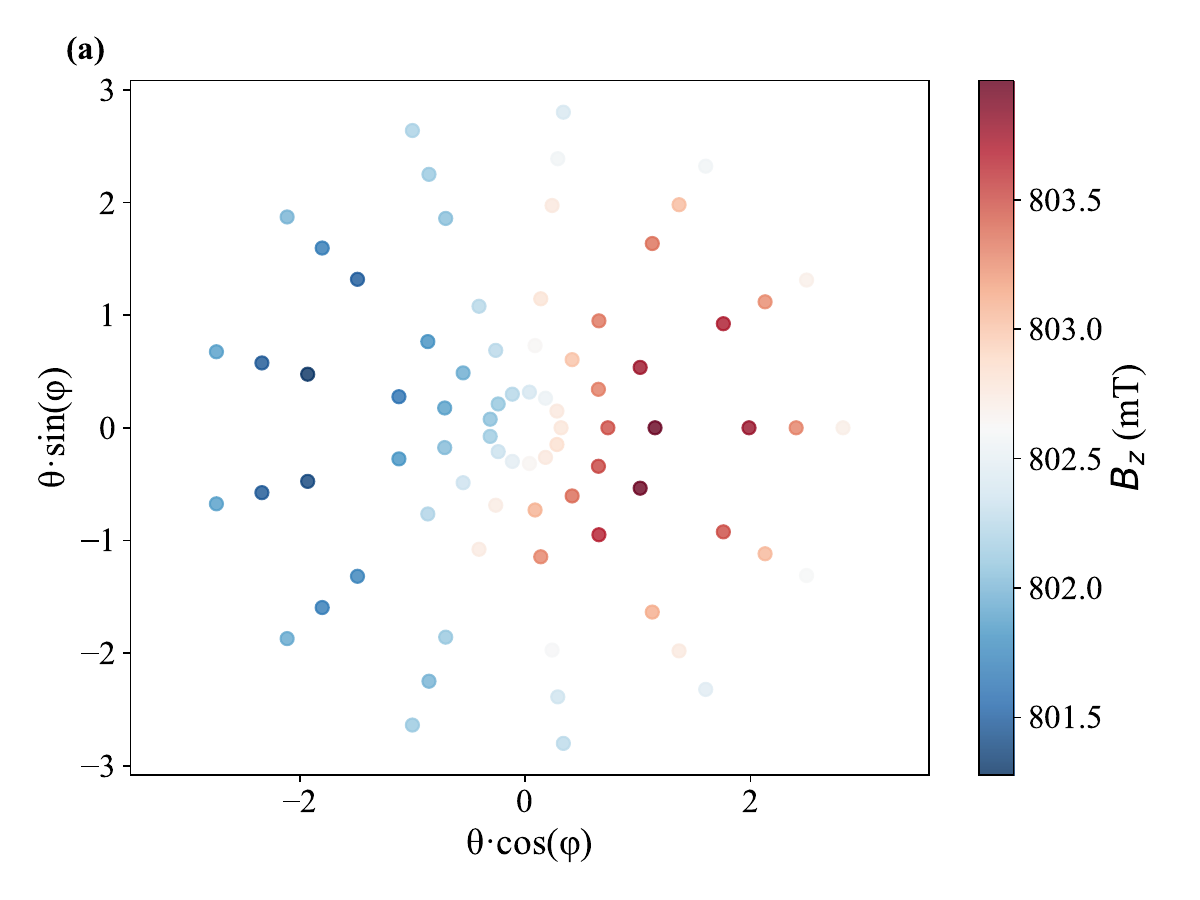} %
\caption*{}
\end{subfigure}
\hfill
\begin{subfigure}{0.48\textwidth}
\includegraphics[width=\linewidth]{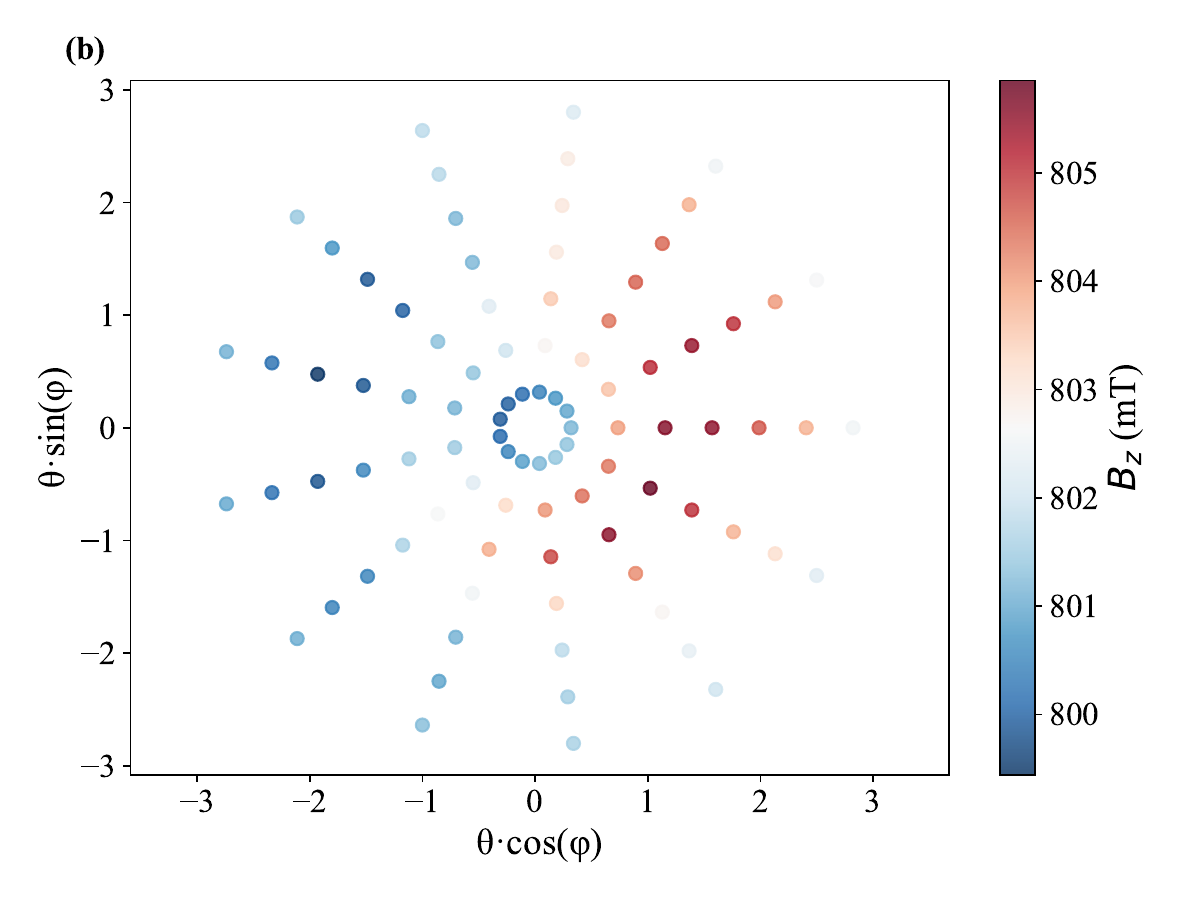} %
\caption*{}
\end{subfigure}
\caption{\label{fig:ball measurements}(a) Magnetic field measurements on the \SI{5}{mm} radius sphere. (b) Magnetic field measurements on the \SI{10}{mm} radius sphere. Data are presented in spherical coordinates with 13 measurement points distributed across each of 7 circular planes.}
\end{figure}

Fig.~\ref{fig:ball measurements} shows the experimental measurement data. We divided a spherical region into 7 circular planes, with 13 measurement points distributed on each plane. Fig.~\ref{fig:ball measurements}(a) displays results for the 5 mm radius sphere, while Fig.~\ref{fig:ball measurements}(b) shows results for the 10 mm radius sphere, both presented in spherical coordinates.

\begin{table*}[!htbp]
\small
\setlength{\tabcolsep}{4pt}
\caption{Spherical harmonic coefficients $a_{lm}$ (in T·m) for expansion orders $l=1$ to $5$. All coefficients are obtained from the TSVD fit with $l_{\max}=5$. Values marked as ``neg.'' are negligible ($<10^{-14}$~T·m).}
\centering
\begin{tabular}{lclclclclc}
\hline\hline
\multicolumn{2}{c}{$l=1$} & \multicolumn{2}{c}{$l=2$} & \multicolumn{2}{c}{$l=3$} & \multicolumn{2}{c}{$l=4$} & \multicolumn{2}{c}{$l=5$} \\
\cline{1-2} \cline{3-4} \cline{5-6} \cline{7-8} \cline{9-10}
Coef. & Val. (T·m) & Coef. & Val. (T·m) & Coef. & Val. (T·m) & Coef. & Val. (T·m) & Coef. & Val. (T·m) \\
\hline
$a_{1,-1}$ & neg. & $a_{2,-2}$ & neg. & $a_{3,-3}$ & neg. & $a_{4,-4}$ & neg. & $a_{5,-5}$ & neg. \\
$a_{1,0}$  & $-1.263\times10^{-2}$ & $a_{2,-1}$ & $1.229\times10^{-6}$ & $a_{3,-2}$ & $-1.358\times10^{-6}$ & $a_{4,-3}$ & $1.169\times10^{-7}$ & $a_{5,-4}$ & $2.588\times10^{-8}$ \\
$a_{1,1}$  & neg. & $a_{2,0}$  & $-2.084\times10^{-6}$ & $a_{3,-1}$ & $1.794\times10^{-6}$ & $a_{4,-2}$ & $3.513\times10^{-7}$ & $a_{5,-3}$ & $1.353\times10^{-8}$ \\
\multicolumn{2}{c}{} & $a_{2,1}$  & $-1.416\times10^{-5}$ & $a_{3,0}$  & $2.233\times10^{-6}$ & $a_{4,-1}$ & $-1.160\times10^{-7}$ & $a_{5,-2}$ & $8.963\times10^{-8}$ \\
\multicolumn{2}{c}{} & $a_{2,2}$  & neg. & $a_{3,1}$  & $5.737\times10^{-7}$ & $a_{4,0}$  & $1.413\times10^{-6}$ & $a_{5,-1}$ & $-7.271\times10^{-8}$ \\
\multicolumn{2}{c}{} & \multicolumn{2}{c}{} & $a_{3,2}$  & $3.314\times10^{-7}$ & $a_{4,1}$  & $2.024\times10^{-7}$ & $a_{5,0}$  & $4.134\times10^{-7}$ \\
\multicolumn{2}{c}{} & \multicolumn{2}{c}{} & $a_{3,3}$  & neg. & $a_{4,2}$  & $2.530\times10^{-7}$ & $a_{5,1}$  & $-3.347\times10^{-8}$ \\
\multicolumn{2}{c}{} & \multicolumn{2}{c}{} & \multicolumn{2}{c}{} & $a_{4,3}$  & $2.444\times10^{-8}$ & $a_{5,2}$  & $1.028\times10^{-7}$ \\
\multicolumn{2}{c}{} & \multicolumn{2}{c}{} & \multicolumn{2}{c}{} & $a_{4,4}$  & neg. & $a_{5,3}$  & $-1.617\times10^{-8}$ \\
\multicolumn{2}{c}{} & \multicolumn{2}{c}{} & \multicolumn{2}{c}{} & \multicolumn{2}{c}{} & $a_{5,4}$  & $1.856\times10^{-8}$ \\
\multicolumn{2}{c}{} & \multicolumn{2}{c}{} & \multicolumn{2}{c}{} & \multicolumn{2}{c}{} & $a_{5,5}$  & neg. \\
\hline\hline
\end{tabular}
\label{tab:coeffs}
\end{table*}

\begin{figure}[!htbp]
\includegraphics[width=0.45\textwidth]{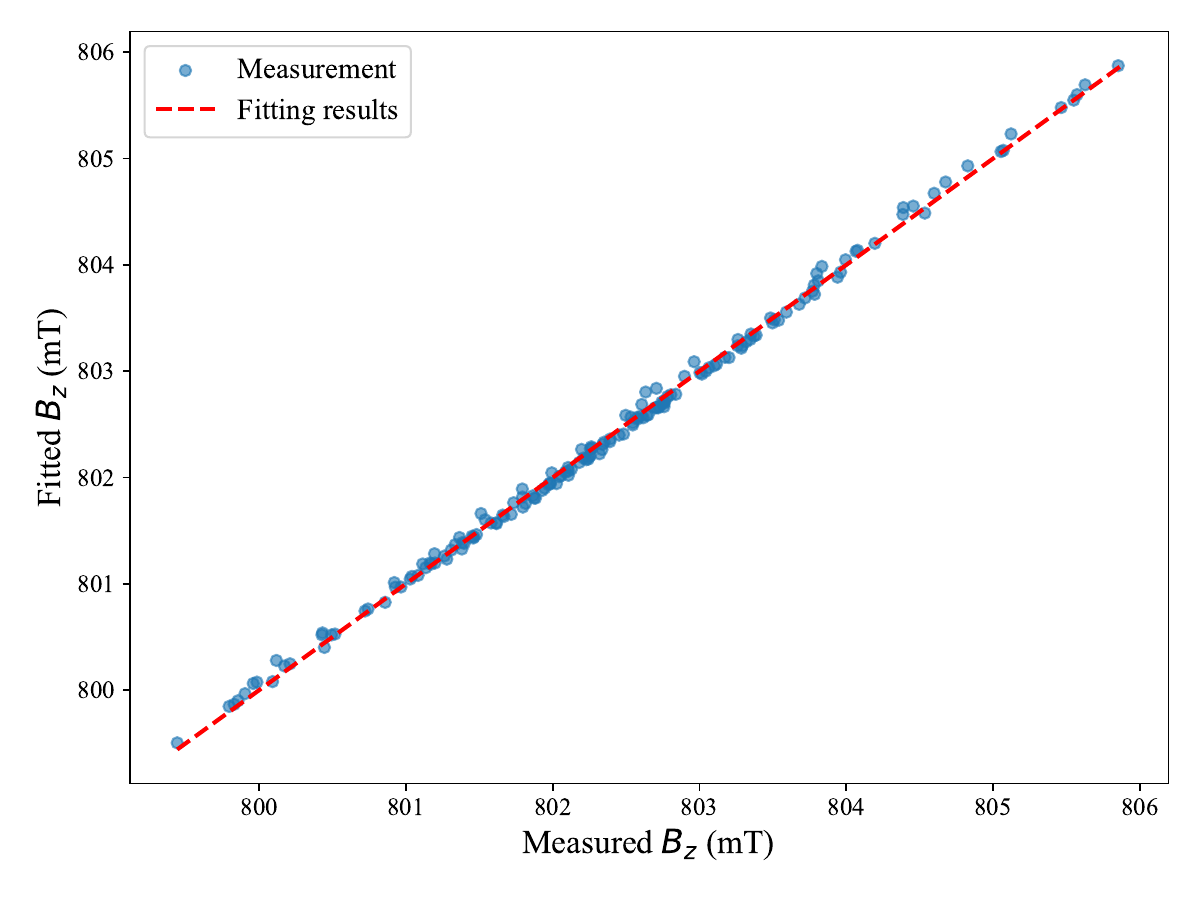} 
\caption{\label{fig:fitting_results} Comparison between measured and fitted $B_z$ for the optimal expansion order $l_{\max}=5$. The root‑mean‑square error is $0.060$\,mT, and the coefficient of determination is $R^2 = 0.9980$, confirming an excellent agreement.}
\end{figure}

For experimental analysis, we employed multi‑sphere joint fitting. Initially, 91 points were measured on each sphere (13 points $\times$ 7 planes). For the 5\,mm radius sphere, 13 points were excluded due to data collection issues, leaving 78 points; for the 10\,mm radius sphere, all 91 points were used.

Fig.~\ref{fig:fitting_results} presents the fitting results for the optimal expansion order $l_{\max}=5$, which was selected after a systematic comparison of $l_{\max}=1$ to $6$. The complete set of 35 spherical
harmonic coefficients is listed in Table~\ref{tab:coeffs}. The coefficient $a_{1,0}=-1.263\times10^{-2}$\,T·m is a dipole coefficient of the scalar potential; it yields the constant central field $B_0$ (zeroth order in space) after taking $B_z = -\partial\Phi/\partial z$. The root‑mean‑square error (RMSE) between the measured and fitted $B_z$ values is $0.060$\,mT and the coefficient of determination is $R^2=0.9980$, confirming an excellent agreement.

To quantitatively justify the choice of $l_{\max}=5$ and to confirm that higher‑order terms are negligible, we compute the maximum contribution of each $(l,m)$ mode on the $5$mm sphere. The mode $(1,0)$ alone contributes $99.67\%$ of the total maximum field corresponding to 802.6 mT. All modes with $l\ge2$ together account for only $0.33\%$, and the contributions of $l=5$ modes are below $0.006\%$. These findings demonstrate that the expansion up to $l_{\max}=5$ captures all physically significant variations; including higher orders would only introduce noise. Therefore, $l_{\max}=5$ is retained as the optimal expansion order.

\begin{figure}[!htbp]
 \begin{subfigure}{0.5\textwidth}
\includegraphics[width=\linewidth]{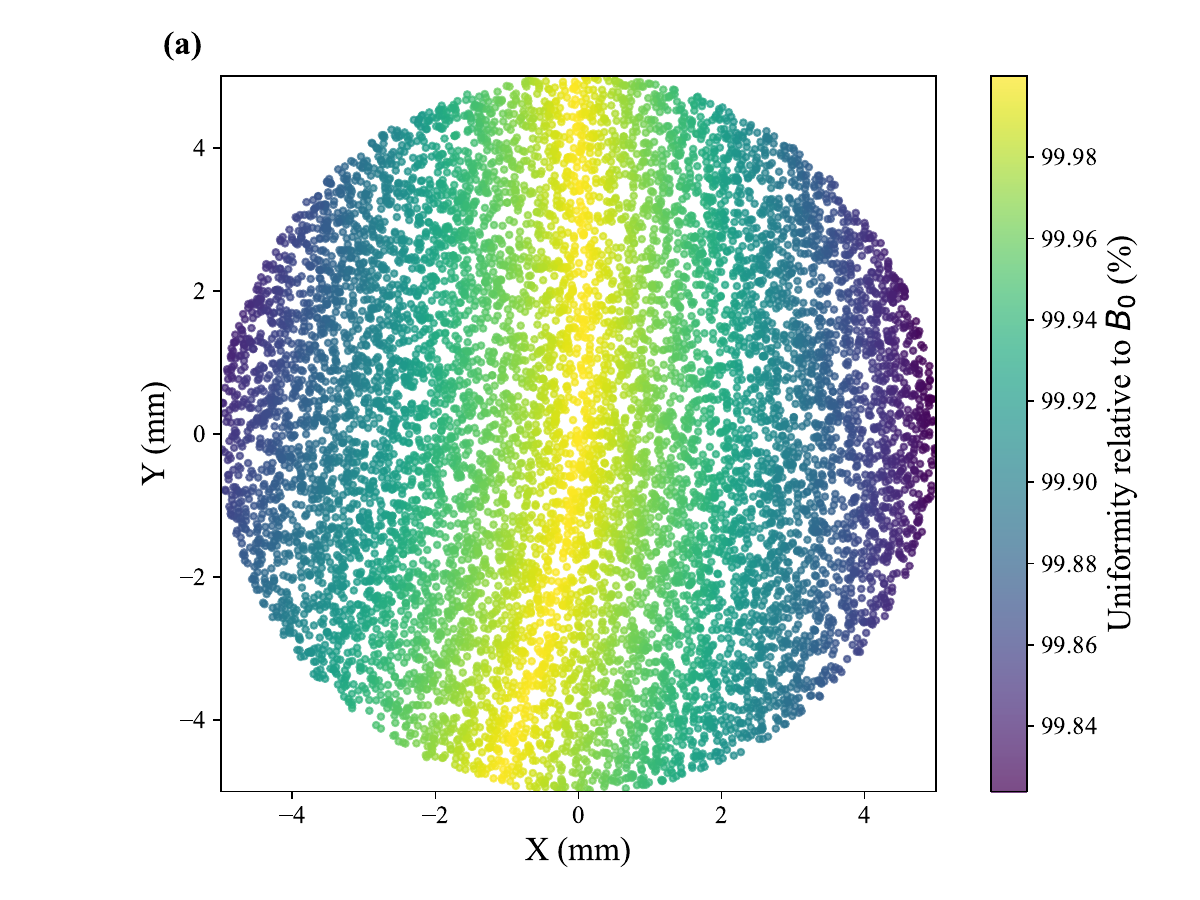} %
\caption*{}
\end{subfigure}
\hfill
\begin{subfigure}{0.5\textwidth}
\includegraphics[width=\linewidth]{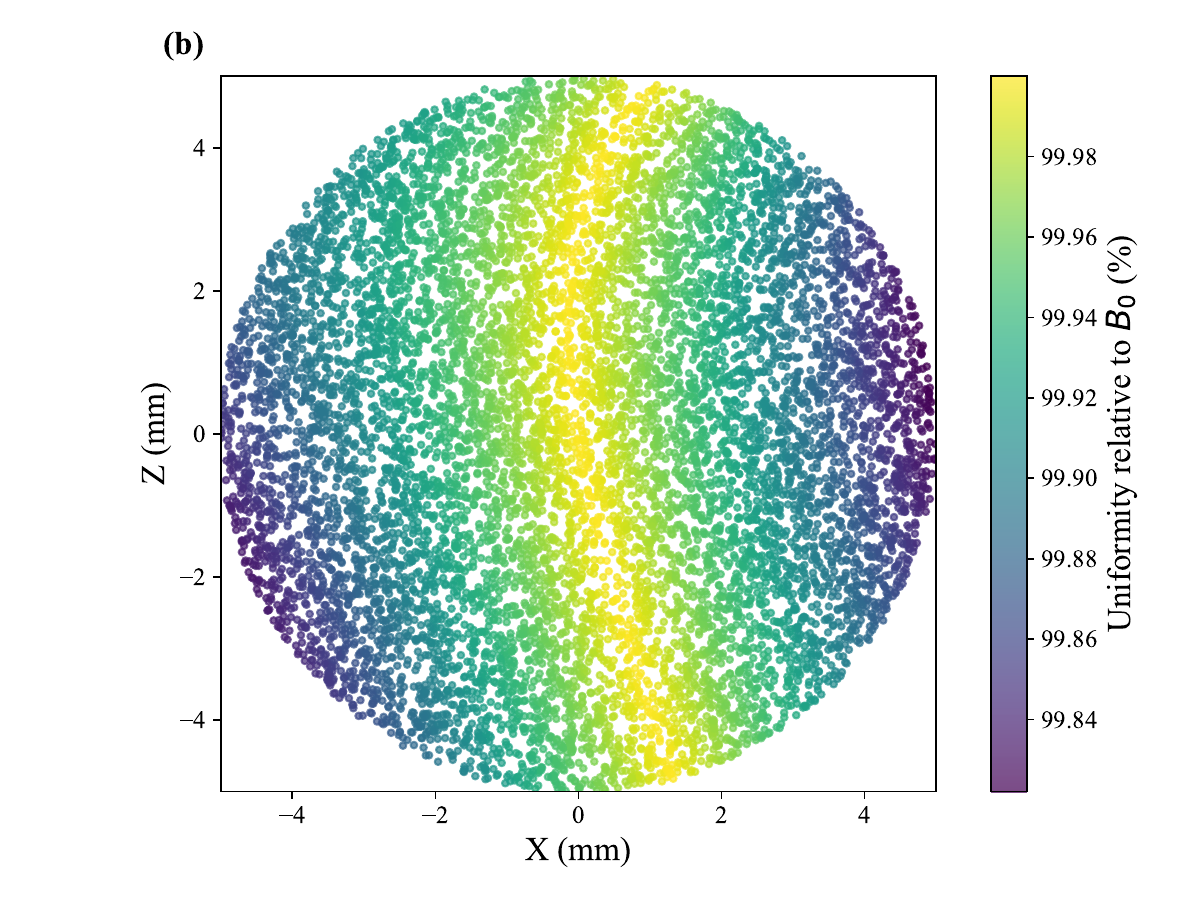} %
\caption*{}
\end{subfigure}
\caption{Field uniformity analysis calculated using Eq.~(3) for uniform regions, with red circles indicating different radii. The color map shows the uniformity relative to $B_{0}$, where $B_{0}$ is the magnetic field strength at center. (a) Field uniformity projected onto the $xy$-plane at $z=0$, displaying the radial dependence. (b) Field uniformity projected onto the $xz$-plane ($y=0$), displaying axial dependence.}
\label{fig:uniformity}
\end{figure}

Using the fitted coefficients for $l_{\max}=5$, the magnetic field was reconstructed throughout the central region via the spherical harmonic expansion. The field uniformity is defined as
\[
\mathrm{Uniformity} = \left(1 - \frac{|B_{z}(\mathbf{r}) - B_{\mathrm{mean}}|}{|B_{\mathrm{mean}}|}\right)\times 100\%, \tag{5}
\]
where $B_{\mathrm{mean}}$ is the average field within the region of interest. For spherical volumes of radii $1$mm, $3$mm and $5$mm, the reconstructed average uniformities (based on the mean field inside each sphere) are $99.988\%$, $99.962\%$ and $99.937\%$, respectively. The remaining non-uniformity is dominated by a dipole term arising in the radial direction. As shown in Fig.~\ref{fig:uniformity}, the transverse gradient in the $xy$-plane is identified with a dipole coefficient $a_{2,1} = -1.42\times10^{-5}$\,T·m, which leads to a discrepancy of $0.03\%$ to $B_0$ at $1$\,mm offset. In contrast, the axial uniformity remains excellent, with a variation of less than $0.01\%$ over $|z| < 1\,\mathrm{mm}$. Higher-order contributions (quadrupole, octupole) are at least four orders of magnitude smaller than the dipole term. This could be due to either the non-uniform remanent magnetization or the magnetic axis displacement.

\section{Conclusion}\label{secIV}

In conclusion, this work presented the design, fabrication and characterization of a compact, high-performance permanent-magnet system based on an optimized stacked structure of fifteen NdFeB ring magnets. The system provides a central magnetic field of 0.8~T within a 44~mm-diameter bore, with a reconstructed field homogeneity of 99.988\% over a spherical volume of 1\,mm radius. This performance is achieved through a tunable NS–SN–NS configuration combined with a symmetric arrangement of the magnets. Spherical harmonic analysis indicates that residual inhomogeneity is dominated by the radial dipole term. With its cryogen-free, power-free operation, compact size, and low cost, the system provides a practical alternative to superconducting magnets for precision ion-trap and FT-ICR mass spectrometry applications. For instance, the compact permanent-magnet FT-ICR system reported by Lemaire et al.~\cite{lemaire2018compact} achieves a mass resolving power of 10,000 using a magnet with a relative homogeneity of 0.5\% over a 3~cm cube (see Fig.~2 of that paper). Because our magnet offers a substantially higher local uniformity (99.937\% over a 1 cm sphere), it is expected to provide an FT-ICR with higher resolution. To further improve the field homogeneity, two achievable methods are recommended. First, either the left or the right magnet group can be replaced so that the two outer groups have closely matched axial profiles, eliminating the intrinsic asymmetry. Second, a couple of room‑temperature shimming coils can be installed to actively compensate residual inhomogeneities. A combination of these measures is expected to increase the local uniformity by at least an order of magnitude. Regarding the magnetic‑axis offset, the ion‑trap cell can be mounted eccentrically to align its axis with the magnetic axis, placing the trap at a location where the transverse gradient field effect can be further reduced.

\textit{Acknowledgement.} The authors extend their sincere gratitude to Dr. Wentian Feng from the Institute of Modern Physics CAS for techniques support. This work was supported by the National Key R\&D Program of  China (No. 2023YFA1606501), the National Natural Science Foundation of China (No. 12474251), Max-Planck Partner Group Project, and the Fudan University Yan Liyuan - EnSiKai Foundation (JX240003).

\bibliography{reference}

\end{document}